\documentclass[12pt]{article}
\usepackage{graphicx}

\newcommand\pubnumber{DPF2013-101}
\newcommand\pubdate{\today}

\def\kennesawtech{${}^a$ Kennesaw State University, Kennesaw, GA 30144, USA\\
${}^b$ Georgia Institute of Technology, Atlanta, GA 30332, USA}
\def\presenter{\footnote{Speaker}}

\def\Title#1{\begin{center} {\Large #1 } \end{center}}
\def\Author#1{\begin{center}{ \sc #1} \end{center}}
\def\Address#1{\begin{center}{ \it #1} \end{center}}

\newcommand\pubblock{\rightline{\begin{tabular}{l} \pubnumber\\
         \pubdate  \end{tabular}}}
\newenvironment{Abstract}{\begin{quotation}  }{\end{quotation}}
\newenvironment{Presented}{\begin{quotation} \begin{center} 
             PRESENTED AT\end{center}\bigskip 
      \begin{center}\begin{large}}{\end{large}\end{center} \end{quotation}}
\def\Acknowledgments{\bigskip  \bigskip \begin{center} \begin{large}
             \bf ACKNOWLEDGMENTS \end{large}\end{center}}

\def\beq{\begin{equation}}
\def\eeq#1{\label{#1}\end{equation}}
\def\eeqn{\end{equation}}

\def\beqa{\begin{eqnarray}}
\def\eeqa#1{\label{#1}\end{eqnarray}}
\def\eeqan{\end{eqnarray}}

\let\bar=\overbar

\def\Dslash{\not{\hbox{\kern-4pt $D$}}}
\def\dslash{\not{\hbox{\kern-2pt $\del$}}}

\def\msb{{\bar{\ssstyle M \kern -1pt S}}}

\begin{document}
\begin{titlepage}
\pubblock

\vfill
\Title{FCNC Top Quark Production via Anomalous Couplings}
\vfill
\Author{Nikolaos Kidonakis$^a$ and Elwin Martin$^b$\presenter}
\Address{\kennesawtech}
\vfill
\begin{Abstract}
We calculate flavor-changing neutral current (FCNC) processes with top-quark 
production via anomalous couplings at LHC energies. 
We report on progress on the FCNC processes $gu \to tZ$, $gu \to t \gamma$ 
and $gu \to tg$. We go beyond leading order and include soft-gluon corrections 
through next-to-next-to-leading order.
\end{Abstract}
\vfill
\begin{Presented}
DPF 2013\\
The Meeting of the American Physical Society\\
Division of Particles and Fields\\
Santa Cruz, California, August 13--17, 2013\\
\end{Presented}
\vfill
\end{titlepage}
\def\thefootnote{\fnsymbol{footnote}}
\setcounter{footnote}{0}

\section{Introduction}
Top quark production can probe various physics beyond the Standard Model,  
including flavor-changing neutral currents (FCNC).  In various new physics 
models such FCNC processes involve anomalous couplings of the top quark.

The effective Lagrangian involving anomalous couplings of a $t,q$ pair 
to electroweak bosons is the following:
\begin{equation}
\Delta {\cal L}_1 =    \frac{1}{ \Lambda } \, e\, 
\kappa_{tqV} \, \bar t \, \sigma_{\mu\nu} \, q \, F^{\mu\nu}_V + h.c.,
\label{tqv}
\end{equation}
where $\Lambda$ is an effective scale which we set equal to the top quark mass,
$e$ is the elementary charge, $\kappa_{tqV}$ is the anomalous FCNC coupling, 
with $q$ a $u$- or $c$-quark and $V$ a photon or $Z$-boson,
$F^{\mu\nu}_V$  are the usual photon/Z-boson field tensors, and 
$\sigma_{\mu \nu}=(i/2)(\gamma_{\mu}\gamma_{\nu}
-\gamma_{\nu}\gamma_{\mu})$ with $\gamma_{\mu}$ the Dirac matrices.
We note that $t$-$u$-$Z$ and $t$-$u$-$\gamma$ couplings numerically dominate the FCNC cross sections, while charm contributions are small due to the small charm parton densities.

Similarly, the Lagrangian that involves anomalous top couplings with gluons is 
\begin{equation}
\Delta {\cal L}_2 =    \frac{1}{ \Lambda } \, g_s\, 
\kappa_{qgt} \, \bar t \, \sigma^{\mu\nu} \, T^a \, \chi \, q \, G^a_{\mu\nu} + h.c.,
\label{tqg}
\end{equation}
where $g_s$ is the strong coupling, $\kappa_{qgt}$ is the anomalous coupling, $T^a$ are the Gell-Mann matrices, $\chi=f^L \, P_L+f^R \, P_R$ with $P_L(P_R)$ the left (right) hand projection operator, and $G^a_{\mu\nu}$  is the gluon field tensor.

Historically, HERA and Tevatron have looked for these processes and set limits on the relevant anomalous couplings.
More recently, the LHC has been used to search for FCNC processes in the 
top-quark sector. ATLAS has set limits of  
$\kappa_{ugt}/\Lambda < 6.9 \times 10^{-3}$ TeV$^{-1}$ and
$\kappa_{cgt}/\Lambda < 1.6 \times 10^{-2}$ TeV$^{-1}$ \cite{ATLAS}.

In this contribution we study higher-order corrections that arise from soft-gluon emission to FCNC top quark processes. In particular we study the processes 
$gu \rightarrow tZ$, $gu \rightarrow t\gamma$, and  $gu \rightarrow tg$. 
The first two have already been studied in some detail in \cite{fcnc} where the one-loop soft-anomalous dimensions were calculated and approximate next-to-leading order (NLO) and next-to-next-to-leading order (NNLO) analytical expressions were derived. Also in \cite{fcnc} the cross sections for FCNC production via $gu \rightarrow tZ$ and $gu \rightarrow t\gamma$ were calculated for Tevatron energy and the higher-order corrections were found to be important. Here we update these calculations for LHC energies. We also consider soft-gluon corrections for the process $gu \rightarrow tg$. For more details on soft-gluon resummation and its applications to Standard Model top quark processes, including top-pair and single-top production, see the review in \cite{review}. 

\section{$gu \rightarrow tZ$}

\begin{figure}[htb]
\centering
\includegraphics[height=1in]{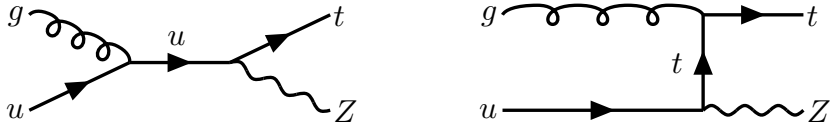}
\caption{Lowest-order Feynman diagrams for $gu \rightarrow tZ$.}
\label{tz}
\end{figure}

We begin with FCNC top production via the process $gu \rightarrow tZ$. The leading-order (LO) diagrams are shown in Fig. \ref{tz}.

\begin{figure}[htb]
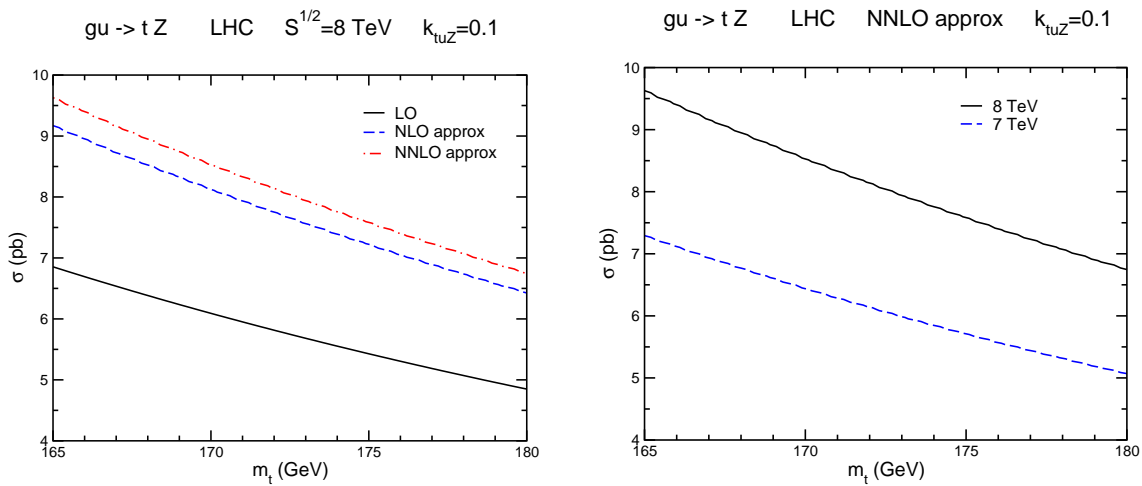

\centering
\includegraphics[height=2.5in]{gutZ8LHCplot.eps}
\hspace{5mm}
\includegraphics[height=2.5in]{gutZNNLOLHCplot.eps}
\caption{(Left) LO, approximate NLO, and approximate NNLO results for $gu \rightarrow tZ$ at 8 TeV LHC energy. (Right) Approximate NNLO results for $gu \rightarrow tZ$ at 7 and 8 TeV LHC energies.}
\label{ZLHC}
\end{figure}

In Fig. \ref{ZLHC} we show results for the cross section for this process at LHC energies as a function of top quark mass, $m_t$. We use the MSTW2008 NNLO parton densities \cite{MSTW} and choose  $\kappa_{tuZ}=0.1$. We also set the factorization and renormalization scales equal to the top quark mass. The left plot shows the LO and approximate NLO and NNLO cross sections at 8 TeV LHC energy. We see that the higher-order soft-gluon corrections significantly increase the cross section. The right plot compares the approximate NNLO results at 7 TeV and 8 TeV energies at the LHC.   

The NLO approximate corrections (i.e. the NLO soft-gluon corrections) provide an around 33\% increase over the LO cross section at 7 and 8 TeV LHC energies.
The sum of the NLO and NNLO approximate corrections provides a total increase of around 40\% over the LO cross section at both 7 and 8 TeV energies, which is very significant. It is interesting to note that the percentage size of the corrections is similar to those at Tevatron energy as calculated in \cite{fcnc}. 

\section{$gu \rightarrow t\gamma$}

\begin{figure}[htb]
\centering
\includegraphics[height=1in]{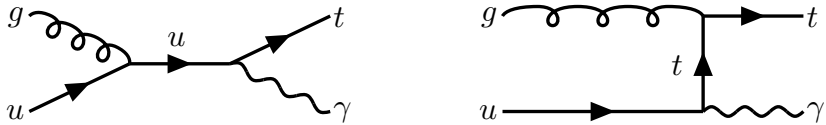}
\caption{Lowest-order Feynman diagrams for $gu \rightarrow t\gamma$.}
\label{tgam}
\end{figure}

We continue with the process $gu \rightarrow t\gamma$. In Fig. \ref{tgam} we show the lowest-order diagrams for this process. Analytical results for the NLO and NNLO soft-gluon corrections were given in \cite{fcnc} where numerical results for Tevatron energy were also shown.

Here we consider this process in $pp$ collisions at the LHC.  
Again, the NLO soft-gluon corrections provide a considerable increase of the LO cross section, around 25\% or more at 7 and 8 TeV LHC energies. However, the NNLO approximate corrections are small (a few percent) and negative, in contrast to $tZ$ production.  

\section{$gu \rightarrow tg$}

Finally, we study the process $gu \rightarrow tg$ which we have not considered before. As mentioned in the Introduction, ATLAS has set limits on the anomalous coupling for this process \cite{ATLAS} and it is important to find how the limits may be affected from changes in the cross section due to higher-order corrections. The color structure of this process is considerably more complicated than for $tZ$ and $t\gamma$ production since here there is a gluon in the final state. In fact the soft anomalous dimensions are $3\times 3$ matrices in color exchange and they are currently being calculated. More results will be reported elsewhere.

\section{Conclusions}

Top quark production via FCNC processes may proceed via various anomalous couplings of the top quark in physics beyond the Standard Model. We have studied several such processes and calculated higher-order corrections through NNLO from soft-gluon emission; such corrections are known to dominate the cross section for top quark Standard Model processes so it is natural to assume the same for FCNC top quark processes. We find that the contributions from these corrections at LHC energies are rather large, up to around 40\% for $gu \rightarrow tZ$ and 25\% for $gu \rightarrow t\gamma$. The process $gu \rightarrow tg$ is theoretically more complicated and is currently under study. 

\Acknowledgments
This material is based upon work supported by the National Science Foundation under Grant No. PHY 1212472.

\end{document}